# Mid-Infrared waveguides and negative refraction with anisotropic metamaterials

Jonathan Plumridge
*Experimental Solid State Group, Physics Department, Imperial College, Prince Consort Road London SW7 2AZ. UK.*

We propose two metamaterial waveguides, operating in the mid-IR, which would display negative refraction. The first waveguide is a metallic strip incorporating quantum wells, whereas the second is a dielectric waveguide which incorporates quantum wells. The negative refraction of both waveguides occurs around the intersubband transition (ISBT) of the quantum wells and is dependent upon the 2D concentration of electron within the wells; these materials could be grown by conventional semiconductor technology (MBE) and the electron concentration within the wells controlled externally by electric fields or optically pumping.

## I. INTRODUCTION

Negative index materials and metamaterials have generated a great deal of interest in the field of classical linear optics. Negative index materials can, in principle, be used to make highly exciting materials such as an invisibility cloak[1] or a perfect lens[2], which have not been practically feasible until the recent development of metamaterials[3]. Now, nano-technology enables one to engineer materials on the length scale much smaller than the wavelength of light; this is a route to controlling the macroscopic electromagnetic properties of a materials response, $\varepsilon$ and $\mu$, and hence a route to making negative index materials.

Negative index and negative refraction are described theoretically by Veselago[4] in 1968, and in this paper Veselago attributes two different, but equivalent physical mechanisms that could lead to a negative index material. Essentially, the Poynting vector and k vector of the travelling wave need to be pointing in opposite directions; this can be achieve by a material possessing negative group velocity and positive k vector (anomalous dispersion[5]), or a material with positive group velocity and negative k vector ($\varepsilon$ and $\mu$ < 0). Causality implies that both mechanisms are highly lossy[4]. The only practical solution to which would be the use of gain to offset the losses or induce anomalous dispersion[6].

So far, the experimental work on negative refraction used anomalous absorption around the plasma frequency of metallic waveguides (an anisotropic slab metallic waveguide[7] and a parallel plate metallic waveguide[8]) to induce negative refraction. In this paper we propose the use of anomalous absorption around the intersubband transition of a quantum well to induce negative refraction, and investigates the use of two types of waveguide: the long range plasmon[9] (LRP section IIb) and a $TM_0$ mode (section IIc). The effects of varying the 2D electron density in the quantum wells are considered, and we show that by increasing the number of electrons in the quantum wells one can induce a negative index region in the dispersion curves of the LRP and $TM_0$ modes whilst maintaining good phase matching conditions. Experimentally the 2D electron density in the quantum wells could be varied by 'pumping' the electrons optically[10], or electrically[11].

Section II provides the theoretical framework for this article. In IIa we present a general description of bound modes supported by an anisotropic layer, IIb outlines the effective medium approach for a LRP waveguide with an embedded quantum well, and IIc details the effective medium approach for a multiple quantum well stack (MQW). In section III is the results section where we present the dispersion curves, absorption, effective refractive index and the effects of field compression of the two designs proposed in section II. Section IV is brief conclusion.

## II. THEORETICAL FRAMEWORK

### A. Bound modes supported by an anisotropic layer

We start by considering the general case of a central homogeneous slab, of thickness $d$ and anisotropic medium of dielectric tensor $\varepsilon_2$, clad by homogeneous isotropic dielectric constant $\varepsilon_1$ (Fig. 1). $\varepsilon_1$ is real and positive while $\varepsilon_2$ is written as the tensor:

$$\bar{\varepsilon}_2 = \begin{pmatrix} \varepsilon_{xx} & 0 & 0 \\ 0 & \varepsilon_{yy} & 0 \\ 0 & 0 & \varepsilon_{zz} \end{pmatrix}, \quad (1)$$

Where $\varepsilon_{xx}=\varepsilon_{yy}$ and $\varepsilon_{zz}$ are complex numbers with as yet no restriction. We follow the procedures outlined by Yeh[12] and solve Maxwell's equations for the boundary condition in Fig.1 for the modes of linearly polarized light which are bound in the z-direction but which propagate in the x-direction. In the case of the transverse magnetic (TM) polarization, the modes at the upper and lower slab surfaces couple symmetrically. The dispersion relation is given by:

$$h \tan(hd/2) = q\varepsilon_{xx}/\varepsilon_1, \quad (2)$$

where $h$ and $q$ are defined by,

$$h^2 = K^2\varepsilon_{xx} - k_x^2\varepsilon_{xx}/\varepsilon_{zz}, \quad (3)$$

$$q^2 = k_x^2 - K^2\varepsilon_1, \quad (4)$$

where $K$ is the free-space wave vector and $k_x$ is the x-component of the wave vector and is known as the complex propagation constant $k_x = k_x^{(r)} - ik_x^{(i)}$. The magnetic fields themselves are given by

$$H_y(x,z,t) = \exp[i(\omega t - k_x x)]H_m(z) \quad (5)$$

$$H_m(z) = \begin{cases} A\cos(hz), & |z| < d/2 \\ B\exp(-qz), & |z| > d/2 \\ B\exp(qz), & |z| < -d/2 \end{cases} \quad (6)$$

In this system we consider that the central slab region ($\varepsilon_2$) is made from repeats of a multi-layered stack, of width $L_s$, and the width of region $\varepsilon_2$ is $N*L_s$, where $N$ is the number of repeats of the multi-layered stack. If $L_s$ is much shorter than the optical wavelength of the mode, then we can use the effective medium approach to estimate the slab's dielectric tensor, $\varepsilon_2$[13].

**B. Effective medium for metallic layer and quantum**

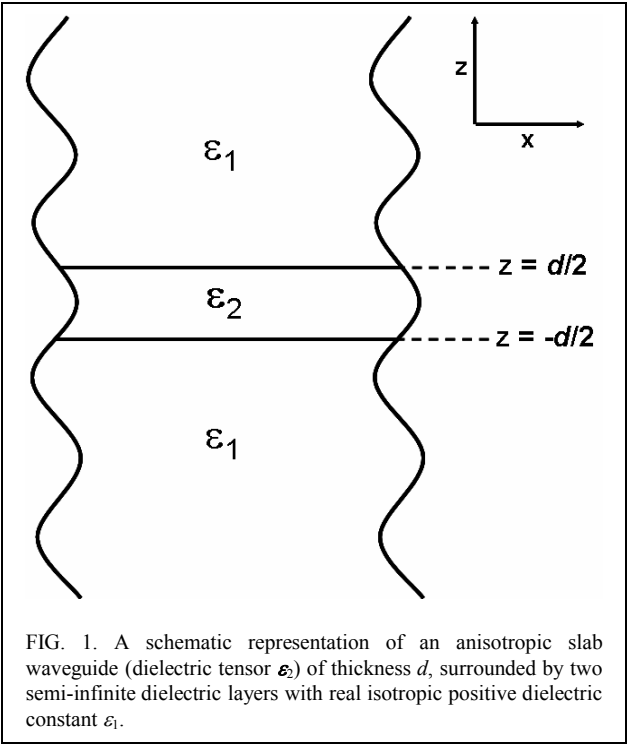

FIG. 1. A schematic representation of an anisotropic slab waveguide (dielectric tensor $\varepsilon_2$) of thickness $d$, surrounded by two semi-infinite dielectric layers with real isotropic positive dielectric constant $\varepsilon_1$.

**well**

In figure 2, we show the multi-layered stack that will be used to form an anisotropic dielectric tensor $\varepsilon_2$; made from a plasmonic metal layer and a single quantum well. The plasmon layer will be a doped semi-conductor whose conductivity is represented by a simple Drude model, whiles the QW layer will be described as a Quasi-two dimensional electron gas[14]. The whole thickness of the multi-layer $L_s$ is much smaller than the wavelength of the guided modes, $\lambda$.

The Fourier components of the electric field, **E**, generalised displacement, **D**, and electron current density, **j**, are connected by the relation

$$\overline{D}(z,k_x;\omega) = \varepsilon(z)\overline{E}(z,k_x;\omega) + \frac{i4\pi}{\omega}\overline{j}(z,k_x;\omega) \quad (7)$$

where $\varepsilon(z)$ is the background dielectric constant at point $z$, and $k_x$ is the parallel component of the photon wavevector. Since we are working in the long wavelength limit, we can assume that $\mathbf{j}(z,k_x;\omega) = \mathbf{j}(z,k_x=0;\omega)$ and leave out references to $k_x$ in our derivation of $\varepsilon_2$.

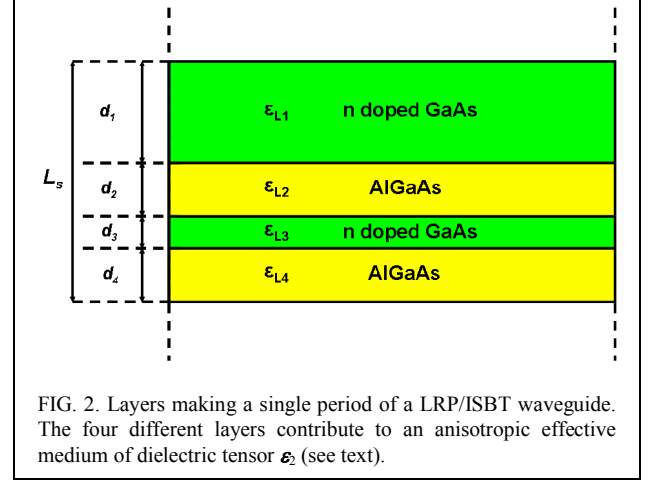

FIG. 2. Layers making a single period of a LRP/ISBT waveguide. The four different layers contribute to an anisotropic effective medium of dielectric tensor $\varepsilon_2$ (see text).

$$\overline{j}(z,\omega) = \overline{E}(z,\omega)\overline{\sigma}(z,\omega) \quad (8)$$

Standard boundary condition require $E_x$ and $D_z$ to be continuous at the interfaces. However, $E_z$ and $D_x$ are not continuous, and it is these quantities which we shall average over to obtain the dielectric response of the effective medium. We define the effective dielectric function as[15]:

$$\varepsilon_{xx}(\omega) = \langle D_x(\omega)\rangle / E_x(\omega), \quad (9)$$

$$\varepsilon_{zz}^{-1}(\omega) = \langle E_z(\omega)\rangle / D_z(\omega). \quad (10)$$

where,

$$\langle D_x(\omega)\rangle = \frac{1}{L_s}\int_{-L_s/2}^{L_s/2} D_x(z,\omega)dz, \quad (11)$$

$$\langle E_z(\omega)\rangle = \frac{1}{L_s}\int_{-L_s/2}^{L_s/2} E_z(z,\omega)dz. \quad (12)$$

There are 4 layers to the multi-layered stack shown in figure 2, and equation 7 relates the background dielectric constant and current density of each layer to its generalised displacement, **D**. The thickness of layer 1 is chosen to be larger than the Fermi wavevector, and hence does not constitute as a quantum well. This would imply that the electrons are free to move as a three dimensional electron gas. However, the thickness of layer 3 is chosen to be short enough to ensure quantum confinement of the electrons in the z-direction, and hence the current density would be described as a quasi-two dimensional electron gas. Layers 2 and 4 have no doping and thus have no associated current density.

For the quasi-two dimensional electron gas (layer 3) we have[13]:

$$\sigma_{zz}^{(2D)} = \frac{1}{E_z(\omega)} \int_{-\infty}^{\infty} j_z(z,\omega) dz = \frac{j_z^{(2D)}(\omega)}{E_z(\omega)}, \quad (13)$$

which describes the non-retarded response of the electron gas to the z-component of the external electric field, **E** (which we will take to be the z-component of the electric field in layer 2; later, we will make use of the continuity of $D_z$ across the boundary from layer 2 to layer 3 in the derivation of $\varepsilon_{zz}$). Which in the local density approximation[16] leads to

$$\sigma_{zz}^{(2D)}(\omega) = \Lambda \frac{-i}{\left[E_{21}^2 - (\hbar\omega)^2\right]/2\hbar\omega\Gamma - i}, \quad (14)$$

where $\Lambda = N_s e^2 f_{21} \hbar / 2m\Gamma$, $E_{21}$ is the intersubband transition energy modified by depolarisation effects, $f_{21} = 2m\hbar^{-2} E_{21}(z_{21})^2$ is the oscillator strength connected with the $1 \rightarrow 2$ transitions, $\tau = \hbar/\Gamma$ is the dephasing time associated with the $1 \rightarrow 2$ transitions, $N_s$ is the surface electron concentration, and finally, $e$ and $m$ are the charge and effective mass of the electron, respectively (restricting ourselves to the two-parabolic-subband model, with only the ground state occupied).

The parallel conductivity of layer 3 is defined by

$$\sigma_{xx}^{(2D)}(\omega) = \frac{1}{E_x(\omega)} \int_{-\infty}^{\infty} j_x(z,\omega) dz = \frac{j_x^{(2D)}(\omega)}{E_x(\omega)}. \quad (15)$$

If we assume a Drude like form then,

$$\sigma_{xx}^{(2D)}(\omega) = \frac{N_s e^2}{m(\tau_\parallel^{-1} - i\omega)}, \quad (16)$$

Where $\tau_\parallel (>\tau)$ is the intrasubband relaxation time.

Similarly, for layer 1 (assuming a Drude like form), we have[17]

$$\sigma_{xx}^{(3D)}(\omega) = \sigma_{zz}^{(3D)}(\omega) = \frac{N_b e^2}{m(\tau_b^{-1} - i\omega)}, \quad (17)$$

where $N_b$ is the bulk 3D electron density and $\tau_b$ is the 3D bulk relaxation time.

Using equations 7 and 11-17 in equations 9 and 10 will yield the following results for the effective medium for the superlattice period shown in figure 2. Note that since $L_s \ll \lambda$, then $E_x$ and $D_z$ will be practically unchanged over the period, $L_s$; $E_x(z,\omega) = E_x(\omega)$ and $D_x(z,\omega) = D_x(\omega)$.

$$\varepsilon_{xx}(\omega) = \widetilde{\varepsilon}_x + \frac{i4\pi}{\omega} \widetilde{\sigma}_{xx}(\omega), \quad (18)$$

$$\widetilde{\varepsilon}_x = \frac{1}{L_s} \sum_i^4 \varepsilon_{Li} d_i, \quad (19)$$

$$\widetilde{\sigma}_{xx}(\omega) = \frac{1}{L_s} \left[\sigma_{xx}^{(3D)}(\omega) d_1 + \sigma_{xx}^{(2D)}(\omega) d_3\right], (20)$$

and

$$\varepsilon_{zz}^{-1}(\omega) = \widetilde{\varepsilon}_z^{-1} - \frac{i4\pi}{\omega} \widetilde{\sigma}_{zz}(\omega), \quad (21)$$

$$\widetilde{\varepsilon}_z^{-1} = \frac{1}{L_s} \sum_i^4 \varepsilon_{Li}^{-1} d_i, \quad (22)$$

$$\widetilde{\sigma}_{zz}(\omega) = \frac{1}{L_s} \left[\frac{\sigma_{zz}^{(3D)}(\omega) d_1}{\varepsilon_{L4}\varepsilon_{L1}} + \frac{\sigma_{zz}^{(2D)}(\omega) d_3}{\varepsilon_{L2}\varepsilon_{L3}}\right]. (23)$$

We have used the continuity of $D_z$ across the boundary of layers 2-3 and layers 1-4 in the derivation of equation 23.

**C. Effective medium approach for multiple quantum wells**

Much work has already been done in describing the optical properties of MQWs in the mir IR. In this paper we adopt the same approach made by Zaluzny and Nalewajko[13] and quote their result from their paper.

$$\varepsilon_{xx}^{mqw}(\omega) = \widetilde{\varepsilon}_x^{mqw} + \frac{i4\pi}{\omega L_{mqw}} \sigma_{xx}^{(2D)}(\omega) L_{qw}, \quad (24)$$

$$\widetilde{\varepsilon}_x^{mqw} = \frac{\varepsilon_b (L_{mqw} - L_{qw})}{L_{mqw}} + \frac{\varepsilon_w L_{qw}}{L_{mqw}}, \quad (25)$$

and

$$\frac{1}{\varepsilon_{zz}^{mqw}(\omega)} = \frac{1}{\widetilde{\varepsilon}_z^{mqw}} - \frac{i4\pi}{\omega L_{mqw} \varepsilon_b \varepsilon_w} \sigma_{zz}^{(2D)}(\omega) L_{qw} \quad (26)$$

$$\frac{1}{\widetilde{\varepsilon}_z^{mqw}} = \frac{L_{mqw} - L_{qw}}{\varepsilon_b L_{mqw}} + \frac{L_{qw}}{\varepsilon_w L_{mqw}}. (27)$$

One should notice that this result is similar to that obtained in II.b.

## III. RESULTS AND DISCUSSION

### A. Propagation, absorption and negative refraction of a metallic and quantum well superlattice waveguide (metallic metamaterial waveguide)

We numerically solve equation 2 for a double superlattice period (FIG. 2.) symmetrically embedded in a dielectric cladding. We will call this waveguide a metallic metamaterial waveguide. The results from which can be used to determine the propagation constant, absorption (and corresponding FOM), group velocity and effective refractive index of the plasmonic mode of the metallic metamaterial waveguide. Here we study two cases: (i) the quantum well has a large surface electron concentration, and (ii) the quantum well has no surface electron concentration. However, in both cases the three dimensional electron concentration in the metallic layer is enough to maintain a long range plasmon mode[9].

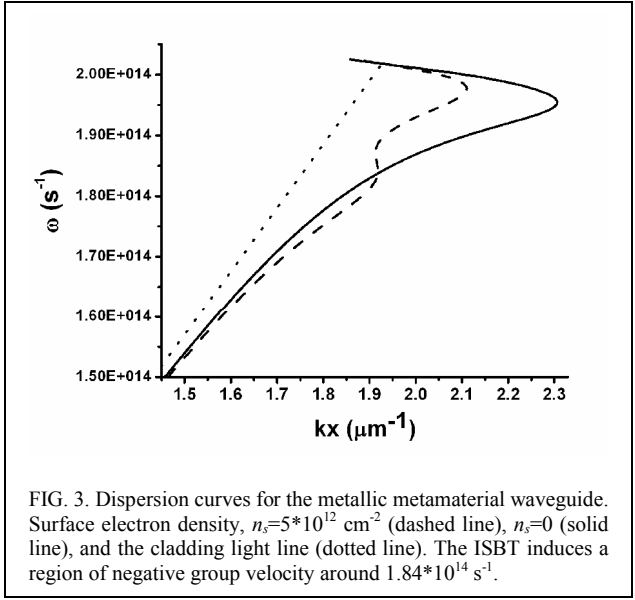

FIG. 3. Dispersion curves for the metallic metamaterial waveguide. Surface electron density, $n_s$=5*10$^{12}$ cm$^{-2}$ (dashed line), $n_s$=0 (solid line), and the cladding light line (dotted line). The ISBT induces a region of negative group velocity around 1.84*10$^{14}$ s$^{-1}$.

The following parameters are used for our supperlattice and cladding layers: $\varepsilon_{cladding}$=8.2067 (Al$_{0.9}$Ga$_{0.1}$As), $\varepsilon_{L1}$=$\varepsilon_{L3}$=10.364 (GaAs), $\varepsilon_{L2}$=$\varepsilon_{L4}$=9.6449 (Al$_{0.3}$Ga$_{0.7}$As), $N_s$=5*10$^{12}$ cm$^{-2}$ (or zero), $m$=0.0665$m_e$, $E_{21}$=123.86 meV, $z_{21}$=2.117 nm, $\Gamma$=5 meV, $N_b$=10$^{19}$ cm$^{-3}$ and we take $\tau_b$=$\tau_{||}$=$\tau$, $d_1$=72 nm, $d_2$=$d_4$=10 nm, and $d_3$=8 nm.

Figure 3 shows the dispersion relation for two superlattice periods, solid line for quantum well doping of $N_s$= 5*10$^{12}$ cm$^{-2}$, and dashed line for quantum well doping of $N_s$=0. The undoped case yield the dispersion curve of a long range plasmon mode[9], and the doped case shows the effects of our anisotropic 'metamaterial', which has a pronounced feature, or kink, around the intersubband energy of the quantum well. From the relation between the propagation constant, phase velocity, $v_p$, and group velocity, $v_g$,

$$v_g(\omega) = \frac{\partial v_p(\omega)}{\partial \omega} = \left(\frac{\partial k_x(\omega)}{\partial \omega}\right)^{-1}. \qquad (28)$$

It is clear, from the dispersion curves in figures 3, that there are regions of positive and negative group velocity. For the undoped quantum well, corresponding to the long range plasmon, the onset of negative group velocity occurs around the bulk plasma frequency of the superlattice – negative refraction from such a waveguide has recently been experimentally verified[7], and is very similar to negative refraction seen in a parallel plate metallic waveguide[8].

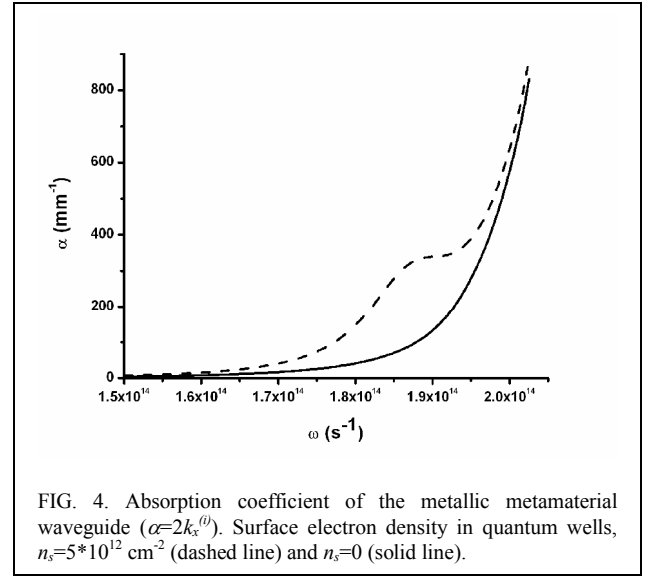

FIG. 4. Absorption coefficient of the metallic metamaterial waveguide ($\alpha$=2$k_x^{(i)}$). Surface electron density in quantum wells, $n_s$=5*10$^{12}$ cm$^{-2}$ (dashed line) and $n_s$=0 (solid line).

However, the purpose of this paper is to draw attention to the negative group velocity induced by the intersubband resonance introduced by the quantum well, furthermore, that the group velocity can be switched from positive to negative values (whilst remaining on the same part of the dispersion curve) by adjusting the number of electron in the quantum well. The 'kink' in the dispersion curve for the doped quantum well superlattice shows exactly this effect occurring where the two dispersion curves cross, at $\omega_I$=1.84*10$^{14}$ s$^{-1}$ ($E_I$=121.27 meV) which is slightly red shifted from the ISBT. At this frequency the 1/e absorption strength of the mode is 2.59*10$^5$ m$^{-1}$, wheras the absorption strength of the undoped quantum well superlattice at this frequency is 6.44*10$^4$ m$^{-1}$ (figure 4). These correspond to FOM values of 14.82 and 59.77.

### B. Propagation, absorption and negative refraction of a multiple quantum well waveguide (MQW waveguide)

The above example uses an ISBT to induce a region of negative group velocity in the dispersion curve of a long range plasmon. The same concept can be extended to a more conventional optical mode, such as a TM mode of a symmetrically clad dielectric slab. In a previous paper we considered how the TM mode of a symmetrically clad MQW can be strongly coupled with the ISBT of that MQW[18]. In this paper, we extend this argument and show at how we can create a region of negative group velocity using the TM mode and ISBT of a MQW slab.

We follow the effective medium approach outlined by reference 18 and derive an anisotropic dielectric tensor, $\varepsilon_2$, for an MQW stack symmetrically surrounded by an isotropic dielectric tensor, $\varepsilon_1$=8.2067 (90% AlGaAs). The parameters for the quantum well and barrier are $\hbar\gamma$= 5 meV, $L_{qw}$= 8 nm, $L_{mqw}$= 20 nm, $E_{21}$=123.86 meV, $z_{21}$=2.117 nm, and $N$=70 (the number of repeats in the MQW stack). For the dispersion curves shown in figure 5, we chose $N_s$ = 0 or $4*10^{11}$ cm$^{-2}$. Figure 5 shows a region of negative group velocity, which for $\omega_l$=1.85*10$^{14}$ s$^{-1}$ ($E_l$= 121.93 meV) yields 1/e absorption strengths of $2.45*10^5$ m$^{-1}$ and 0 m$^{-1}$ for the doped and undoped MQWs respectively (these correspond to FOM values of 14.66 and ∞), figure 6.

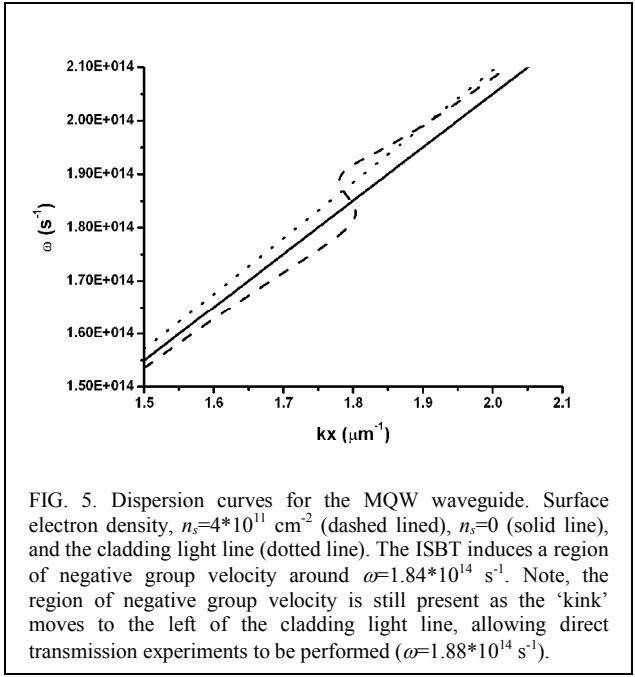

FIG. 5. Dispersion curves for the MQW waveguide. Surface electron density, $n_s$=4*10$^{11}$ cm$^{-2}$ (dashed lined), $n_s$=0 (solid line), and the cladding light line (dotted line). The ISBT induces a region of negative group velocity around $\omega$=1.84*10$^{14}$ s$^{-1}$. Note, the region of negative group velocity is still present as the 'kink' moves to the left of the cladding light line, allowing direct transmission experiments to be performed ($\omega$=1.88*10$^{14}$ s$^{-1}$).

It is useful to note that part of the dispersion curve in FIG. 6. for $n_s$=4*10$^{11}$ cm$^{-2}$, crosses over to the left hand side of the cladding light line. There is thus a region of negative index material which is accessible without the need of a prism coupling or diffraction coupling scheme. This would allow negative refraction experiments from direct transmission to be performed, as was done in reference 8.

## C. Negative Refraction and effective negative refractive index

The ideas of group and phase velocity can be used to define an effective refractive index for light propagating in the mode[8],

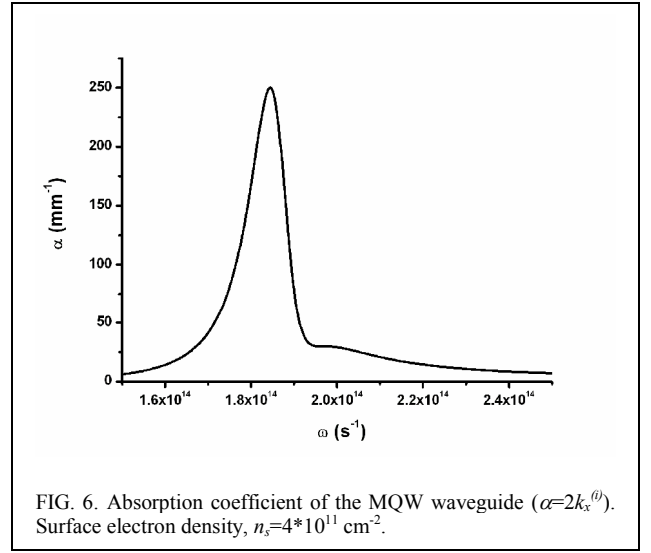

FIG. 6. Absorption coefficient of the MQW waveguide ($\alpha$=2$k_x^{(i)}$). Surface electron density, $n_s$=4*10$^{11}$ cm$^{-2}$.

$$n_{mode} = \text{sgn}(\bar{v}_p \bullet \bar{v}_g)c/|\bar{v}_p|, \quad (29)$$

which, at $\omega$=1.84*10$^{14}$ s$^{-1}$, flips sign from 3.146 to -3.121 for the metallic metamaterial waveguide, and 2.918 to -2.904 for MQW waveguide. This would yield a dramatic and almost instantaneous change in the sign of the angle of refraction, which could in principle be optically switched, either by optically injecting electrons from the valence band into the conduction band of the quantum well[10], or by designing a 3 level quantum well with levels 2-3 providing the region of negative refractive index (level 2 being optically pumped from level 1).

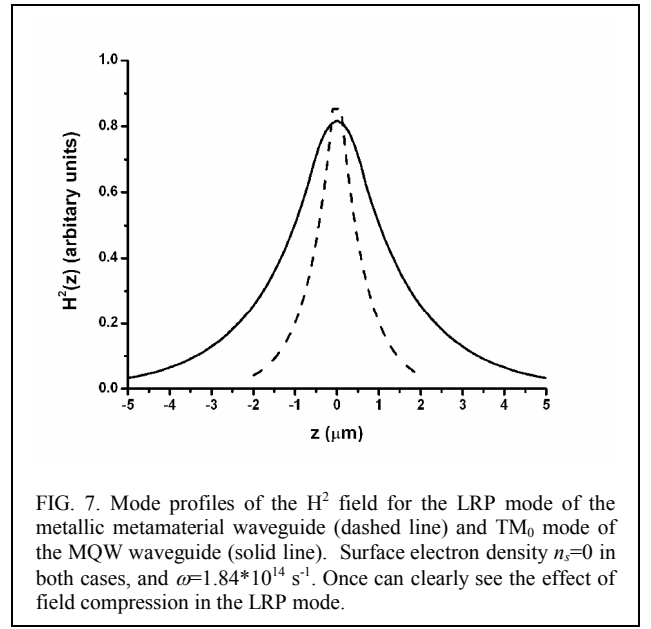

FIG. 7. Mode profiles of the H$^2$ field for the LRP mode of the metallic metamaterial waveguide (dashed line) and TM$_0$ mode of the MQW waveguide (solid line). Surface electron density $n_s$=0 in both cases, and $\omega$=1.84*10$^{14}$ s$^{-1}$. Once can clearly see the effect of field compression in the LRP mode.

## D. Field enhancement and compression

In figure 3, the bare LRP mode is much further to the right of the cladding light line than the bare TM$_0$ mode in figure 5; this effect is more pronounced as one tunes closer to the plasma frequency of the LRP mode. As a consequence there is more potential for sub-wavelength field compression for the sample A. Figure 7 shows the reconstructed fields of the bare LRP and $TM_0$ modes at $\omega=1.84*10^{14}$ s$^{-1}$ for comparison ($n_s=0$ in both cases).

## IV. CONCLUDING REMARKS

In conclusion, we have proposed two metamaterial waveguides that use the ISBT of incorporated quantum well to induce a region of negative group velocity (which is equivalent to a negative refractive index). The effect is strongly dependent upon the 2D electron density in the quantum well, and as such, one should be able to pump electron into the wells (optically or electrically) and thereby make the waveguide into an active component. The predicted losses in the negative index waveguides are quite small compared to other proposed schemes, as can be seen by the high FOM values. Also, more complicated quantum well structures could be used to induce gain (or even gain without population inversion[19]) into the medium, thereby overcoming the losses.

### ACKNOWLEDGEMENTS

The author would like to Robert Steed for many stimulating discussions. Funding from the UK Engineering and Physical Sciences Research Council is gratefully acknowledged.